\documentclass[aps,pre,reprint,bibnotes,footinbib,showkeys,groupedaddress,superscriptaddress]{revtex4-1}
\pdfoutput=1
\usepackage[utf8]{inputenc}
\usepackage{bm,amsmath,amssymb,mathtools,graphicx}
\usepackage{epic}
\usepackage{natbib}
\usepackage{xcolor}
\usepackage[colorlinks=true,urlcolor=blue,citecolor=blue,linkcolor=red,hyperfootnotes=false]{hyperref}
\usepackage{textcomp,gensymb,units,grffile}
\usepackage[a4paper]{geometry}
\usepackage[normalem]{ulem}

\newcommand{\be}{\begin{equation}}
\newcommand{\ee}{\end{equation}}
\newcommand{\bea}{\begin{eqnarray}}
\newcommand{\eea}{\end{eqnarray}}
\newcommand{\nn}{\nonumber}
\newcommand{\ba}{\begin{align}}
\newcommand{\ea}{\end{align}}
\newcommand{\y}{\textbf{y}}
\newcommand{\T}[1]{\text{#1}}
\newcommand{\tl}[1]{\tilde{#1}}
\newcommand{\rl}{\rho_b^{\text{\T{lin}}}}
\newcommand{\rt}{\rho_b^{\text{\T{tree}}}}
\newcommand{\plin}{\mathcal{P}_B^{\text{\T{lin}}}}
\newcommand{\plim}{\mathcal{P}_{mB}^{\text{\T{lin}}}}
\newcommand{\ppop}{\mathcal{P}_B^{\text{\T{tree}}}}
\newcommand{\ppo}{\mathcal{P}^{\text{\T{tree}}}}
\newcommand{\pli}{\mathcal{P}^{\text{\T{lin}}}}
\newcommand{\trans}{\mathbb{T}}
\graphicspath{{Figures/}}

\begin{document}

\title{Linking lineage and population observables 
in biological branching processes}
\author{Reinaldo  Garc\'{i}a-Garc\'{i}a}
\author{Arthur Genthon}
\author{David Lacoste}

\email{reinaldomeister@gmail.com}

\affiliation{Gulliver, ESPCI Paris, PSL University, CNRS, 75005 Paris, France}

\begin{abstract}
Using a population dynamics inspired by an ensemble of growing cells, 
a set of fluctuation theorems linking observables measured
at the lineage and population levels are derived. 
One of these relations implies specific inequalities comparing  
the population doubling time with the mean generation time 
at the lineage or population levels. 
While these inequalities have been derived before 
for age controlled models with negligible mother-daughter correlations,
we show that they also hold for a broad class of size-controlled models.
We discuss the implications of this result for the interpretation of a recent
experiment in which the growth of bacteria strains has been probed at the single cell
level.
\end{abstract}

\maketitle

\section{Introduction}

The question of how a cell controls its size is a very old one \cite{Koch2001}, 
which despite decades of research is 
still under intense focus, because the old experiments have only provided incomplete answers 
while a new generation of experiments based on the observation and manipulation 
of single cells in microfluidic devices is becoming more and more mature \cite{Willis2017}.
For instance, with time-lapse single cell video-microscopy, entire lineages of single cells 
such as {\it E. coli} can be traced over many generations.
These experiments allow to investigate mechanisms of cell size control (cell size homeostasis) 
with unprecedented statistics both at the single cell level and at the level of a population.

Many policies of cell size control have been introduced: the ``sizer'' in 
which the cell divides when it reaches a certain size, the ``timer'' in which
the cells grows for a specific amount of time before division, and the 
``adder'' in which cells add a constant volume each generation \cite{Amir2014}. 
The adder principle is now favored 
by many experiments \cite{Robert2014,Taheri-Araghi2015,Grilli2017,Campos}, 
yet there is no consensus on why a specific regulation emerges under certain conditions, 
and how it is implemented at the molecular level. 

Another important question is how to relate measurements made at the lineage and 
at the population levels. 
A classical study revealed the discrepancy between the mean generation time and 
the population doubling time \citep{Powell1956} in an age-dependent branching process with no 
mother-daughter correlations,  
called Bellmann-Harris process in the literature on branching processes \cite{Kimmel2015}. 
Importantly, it is still not known at present how to relate the mean generation time and the population 
doubling time in general models of cell size control.

Inspired by single-cell experiments with colonies of prokaryotic cells 
in microfluidic devices \cite{Hashimotoa2016,Taheri-Araghi2015}, 
we consider here continuous rate models (CRM), based 
on stochastic differential equations \cite{Hall1991,Robert2014}. 
The population dynamics generated by CSM has an interesting thermodynamic structure 
uncovered in Refs.~\cite{Kobayashi2015,Sughiyama2015}, which we also 
exploit here to derive new fluctuation relations. 
As usual with fluctuation theorems \citep{Crooks2000_vol61},   
our results map typical behaviors in one ensemble (here the population level) 
to atypical behaviors in another one (here the single lineage level).
A similar connection lies at the basis of an algorithm to measure large deviation 
functions using a population dynamics \cite{Nemoto2016,Giardina2006}.
In the mathematical literature on branching processes, relations of this kind
are known as Many-to-One formulas \cite{Bansaye2011}; they explain the existence of a statistical biais, when 
choosing uniformly one individual in a population as opposed to following a lineage.

This paper is organized as follows: In the section \ref{CRM}, 
we introduce two different averaging procedures for CRM dynamics, 
which we call the tree and the lineage averages. 
In section \ref{sec:size}, we define and study size-controlled models. 
This includes a derivation of a fluctuation relation in terms of a quantity which we 
call dynamical activity. Such a fluctuation relation maps the single lineage level 
and the population level. We test it numerically, and we show that it 
can be used to determine the population growth rate from lineage statistics. 
In section Sec.~\ref{size-2}, we derive a second more general fluctuation relation. 
We explain why this result is related to the notion of ``fitness landscape'' 
introduced in Ref.~\cite{Nozoe2017}, and we derive from it important inequalities
comparing the mean generation times at the lineage and tree levels
with the population doubling time. We then discuss the implications of our results for 
the experiment carried out by Hashimoto et al. \cite{Hashimotoa2016}, in which 
these inequalities have been tested.
Then, we analyze age models with and without correlations 
between mother and daughter cells in Sec.~\ref{sec:age}. 
Finally, we conclude in Sec.~\ref{conc}, 
while some important technical details are given in the appendices.

\section{Tree and lineage averages in a population}
\label{CRM}

\begin{figure}[t]
\centering
  \includegraphics[scale=0.38]{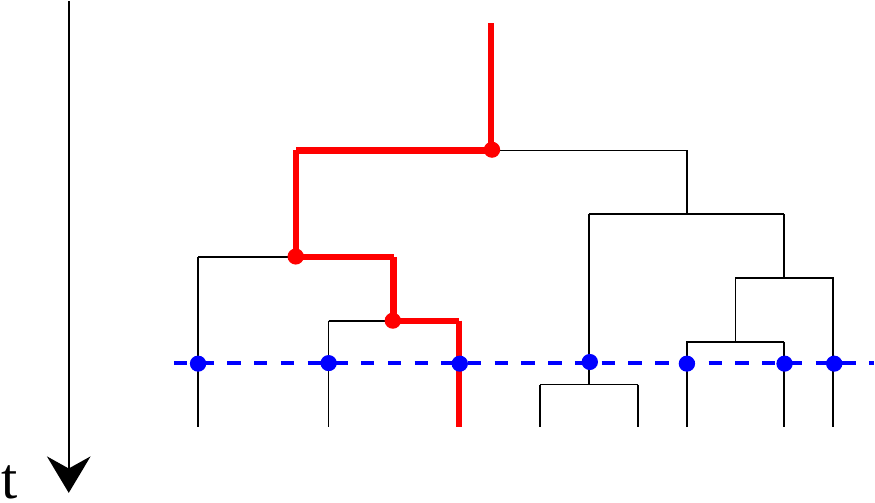}
  \caption{Representation of the three main levels of description of the ensemble of cells: 
the lineage level (bold, red), the population snapshot (horizontal, dashed, blue) and the entire tree (thin, black).
A lineage average is an average over a single lineage going forward in time from 
the ancestor of the colony to its final state at time $t$. A tree average is equivalent to the average over \emph{all} lineages starting at time $t$ from those cells currently alive and going backwards in time up to the ancestor of the colony.}  
\label{fig1}
\end{figure}

Let us consider a population of cells as shown in Fig.~\ref{fig1}, which grow by division 
into only two offsprings at the end of each cell cycle.
This population dynamics can be studied at three distinct levels : the lineage
level (red), the population snapshot (blue) and the tree level which includes 
the complete phylogeny \cite{Lin2017}.

In the following, we shall introduce two different averages corresponding to the 
tree and lineage levels, which are defined for a fixed initial time $t=0$ 
and final time $t$ of the dynamics of the population.
For a tree average, we consider all the branches of the tree including 
all the cells still present at time $t$ and which will divide only after the time $t$.
This tree distribution puts an equal weight on the lineages which end at time $t$ and 
from that point, it goes backward in time towards the original ancestor 
from which the population originated.
For this reason, this distribution is known in the literature under the name of 
retrospective distribution \cite{Sughiyama2015} and is equivalently 
an average over histories, {\it i.e.} backward lineages \cite{Thomas2017}.
In contrast to this, what we call in this paper a lineage average, corresponds to an 
average done over forward lineages, which go forward in time from the original ancestor 
of the colony towards its final state at time $t$.

\begin{figure}[t]
\centering
  \includegraphics[scale=0.39]{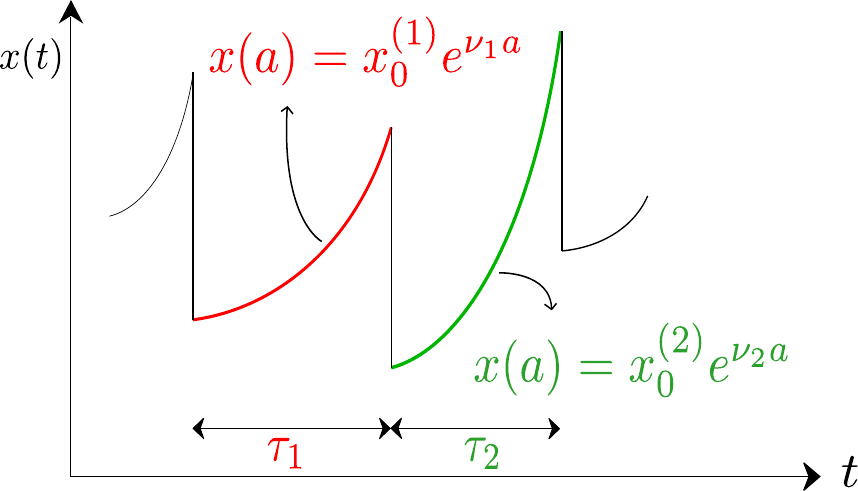}
  \caption{Evolution of the cell size $x(t)$ along a lineage. The cell cycle $i$ is parametrized by 
three random variables: the generation time $\tau_i$, the growth rate $\nu_i$ and the size at birth $x_0^i$.}
  \label{fig1b}
\end{figure}

For bacteria such as E. coli growing in a rich medium, each cell cycle is well
described by an exponential growth phase \cite{Brenner2015}, which for the cell cycle $i$ can be 
parametrized by only three random variables shown in Fig.~\ref{fig1b}: the size at birth $x_0^i$, the  
growth rate $\nu^i$ and the generation time $\tau_i$.

\section{Results for size-controlled models}
\label{sec:size}
\subsection{Definition of size-controlled models}

Let us first consider a model with size-dependent division rate.
The evolution of the number of cells of size $x$ and single cell growth rate $\nu$
at time $t$, $n(\y,t)$ with $\y=(x,\nu)$, obeys the equation \cite{Hall1991,Robert2014}:
\bea
\label{CRMpopulation2}
\partial_t n(\y,t) &=& -\nu\partial_x [x n(\y,t)]-B(\y) n(\y,t) \\ \nn
&+& 2 \int d\y' \Sigma 
(\y | \y') B(\y') n(\y',t),
\eea
where $B(\y)$ is the division rate and $\Sigma(\y | \y')$ is the  
probability for a newborn cell to have parameters $\y$ given that 
the mother cell has parameters $\y'$. By integrating Eq.~\eqref{CRMpopulation2} 
over $\y$ using the condition $\int d\y \Sigma(\y | \y')=1$, a  
deterministic equation of evolution of the total population 
$N(t)=\int d \y n(\y,t)$ is obtained.
The instantaneous growth rate of the population is defined as 
$\Lambda_p(t)=\dot{N}/N$, while the growth rate of the total 
volume of the cells is $\Lambda_V(t)=\dot{V}/V$ with $V(t)=\int d\y x n(\y,t)$.
When a steady state for the variable $\y$ is reached, both $\Lambda_p$ 
and $\Lambda_V$ become independent of time and equal to each other \cite{Lin2017}.

If instead of the full population, we consider the dynamics at the lineage level, the natural 
quantity to study is the \emph{probability density} of the cell to have
size $x$ and growth rate $\nu$ at time $t$, $p(x,\nu,t)$, which satisfies the evolution equation
\bea
\label{CRMlineage2}
\partial_t p(\y,t) &=& -\nu\partial_x [x p(\y,t)]-B(\y) p(\y,t) \\ \nn
&+& \int d\y' \Sigma 
(\y | \y') B(\y') p(\y',t).
\eea
Note the difference with Eq.~\eqref{CRMpopulation2} due to the absence of the factor
 $2$ in front of the integral, rendering $p(\y,t)$ normalizable at any time, $\int\, p(\y,t)d\y=1$.

\subsection{Fluctuation theorem for dynamical activity}
\label{size-1}

We now address the problem of connecting lineage to  
tree or population snapshot statistics in models with size control.
The evolution of a given cell from time 0 to the time $t$  
is encoded in the trajectory $\{ \y \}_0^t=\{ \y \}$. 

For the case of size-controlled model, we derive 
in Appendices~\ref{app:path},  
path probabilities representations at the population and lineage levels, which are given by 
\eqref{path-lin} and~\eqref{path-pop-2} respectively.
Comparing these two expressions, we see that a possible way to bring both distributions 
``closer'' together, is to multiply the division rate at the lineage level by the factor $m$, 
and to consider a lineage starting from the same initial condition as that of the population. 

Then, we introduce the dynamical activity
$W_t( \{\y \}) = \int_0^t dt' B(\y(t'))$, which quantifies the activity of  
cell divisions, and the time averaged population growth rate 
\be
\label{GR}
\Lambda_t=\frac{1}{t} \int_0^t dt' \Lambda_p(t')=\frac{1}{t} \ln \frac{N(t)}{N(0)}.
\ee

After multiplying the relation mentioned above between path probabilities by an arbitrary trajectory-dependent 
observable $A(\{ x,\nu \})$, and after taking the average, for the special case where $m=2$,
one obtains the following fluctuation relation :
\be
\label{FT}
\langle A( \{ \y \} ) \rangle_{\T{tree},B}= \langle A( \{ \y \} )  
e^{W_t( \{\y \} )-t \Lambda_t} \rangle_{\T{lin},2B},
\ee
where $\langle .. \rangle_{\T{tree},B}$ denotes a tree average generated by the original 
dynamics with a division rate $B$ while $\langle .. \rangle_{\T{lin},2B}$ denotes a 
lineage average with a modified dynamics that has a division rate $2B$. 
The reason for this modified division rate is that each cell divides
into $m=2$ cells, as a result a factor two appears at the population level in Eq.~\eqref{CRMpopulation2},
which is absent for the corresponding equation at the lineage level.
In the particular case where the observable $A$ only depends on 
$\y(t)$ instead of the full trajectory $\{ \y \}$, 
Eq.~\eqref{FT} relates the lineage level to the population snapshot level 
instead of the tree level.
The mapping also requires that the original and the modified dynamics
start with the same initial condition $\y(0)$, defined here 
in terms of cell size and growth rate.

\begin{figure}[t]
\centering
  \includegraphics[scale=0.32]{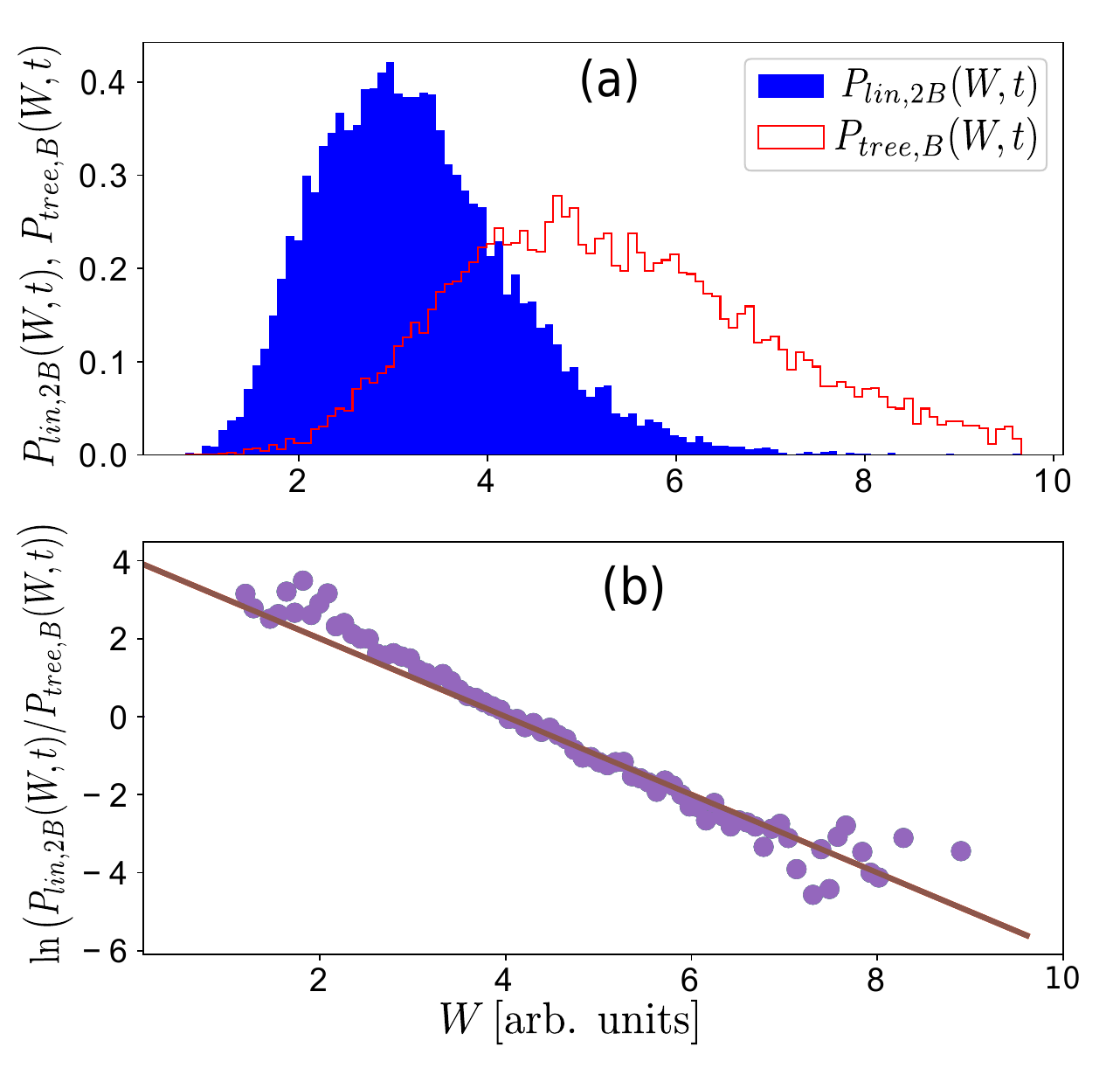}
  \caption{Illustration of the fluctuation relation in the case of growth 
with a constant growth rate $\nu=2$. (a) Distributions 
of dynamical activity at the time $t=2$ in a lineage with division rate $2B$ (blue, filled) and in a tree with division rate $B$(red, unfilled). (b) 
Log-ratio of these probability distributions.}
  \label{fig2}
\end{figure}

For the specific choice $A( \{ \y \} )=\delta (W - W_t( \{\y \})$, Eq.~\eqref{FT} 
leads to Crooks-like relation \cite{Crooks2000_vol61}:
\be
\label{crooks}
P_{\T{tree},B}(W,t)=P_{\T{lin},2B}(W,t) e^{W - t \Lambda_t},
\ee
which relates the distribution of dynamical activity at time $t$ 
in a tree (resp. in a lineage): 
$P_{\T{tree},B}(W,t)$ (resp. $P_{\T{lin},2B}(W,t)$).
This relation is illustrated in Fig.~\ref{fig2} for a 
population of cells growing with a constant single cell growth rate $\nu$.
Numerically, instead of working directly with Eq.~\eqref{CRMpopulation2},
we simulate an equivalent Langevin equation, which accounts 
for deterministic growth with the rate $\nu$ and 
stochastic cell divisions with a rate $B(x,\nu)$. 
In the simulation, the division has been assumed to be symmetric
and the single cell growth $\nu$ constant,
which corresponds to the particular choice of 
$\Sigma(\y | \y')=\delta(\nu-\nu') \delta(x-x'/2)$.
Note that this dynamics bears some similarity to that of   
stochastic resetting introduced in Ref. \cite{Evans2011},  
with the difference that in our case the resetting of the size is relative to the current 
size before division, while in this reference the resetting was to a constant position.
Another important difference is the absence of diffusion in our model.

We have used normalized units of time and size, so that 
$\nu=2$ and $B(x,\nu)=\nu x$ in these units.
Since $\Lambda_p=\Lambda_V=\nu$, $\Lambda_t=2$, the two distributions 
measured at the time $t=2$ cross as expected at $W=4$ (Fig.~\ref{fig2}(a)). 
Fig.~\ref{fig2}(b) confirms that the slope of the log-ratio of the 
two probability distributions is indeed $-1$ as expected from Eq.~\eqref{crooks}. 

Let us emphasize the following points concerning our first main result:
This fluctuation relation is very general, it holds whether or not the single cell 
growth rate fluctuates, i.e., for arbitrary forms of the kernel $\Sigma$ and 
arbitrary division rate $B(x,\nu)$. 
There is no requirement that the population should be stationary neither at time $0$ 
nor at time $t$. Further, it generalizes to the case that each cell 
has $m$ offsprings instead of two, provided that 
this number $m$ is independent on the state of the system $\y$ and that the 
modified lineage dynamics has a division rate $mB(x,\nu)$, as shown in Appendix~\ref{app:FT-size}. 

The normalization of $P_{\T{tree},B}(W,t)$ in~\eqref{crooks} leads to the relation:
\be
\label{FT_Lambda}
\Lambda_t=\frac{1}{t} \ln \int dW e^W P_{\T{lin},2B}(W,t),
\ee
which could be used either to infer the population growth rate from lineage trajectories or
to infer the form of the division rate $B$ using lineage and population trajectories \cite{Doumic2009}.
In the next subsection below, we provide such a numerical illustration.

\subsection{Application to the determination of a population growth rate}

\begin{figure}[t]
\centering
\includegraphics[scale=0.355]{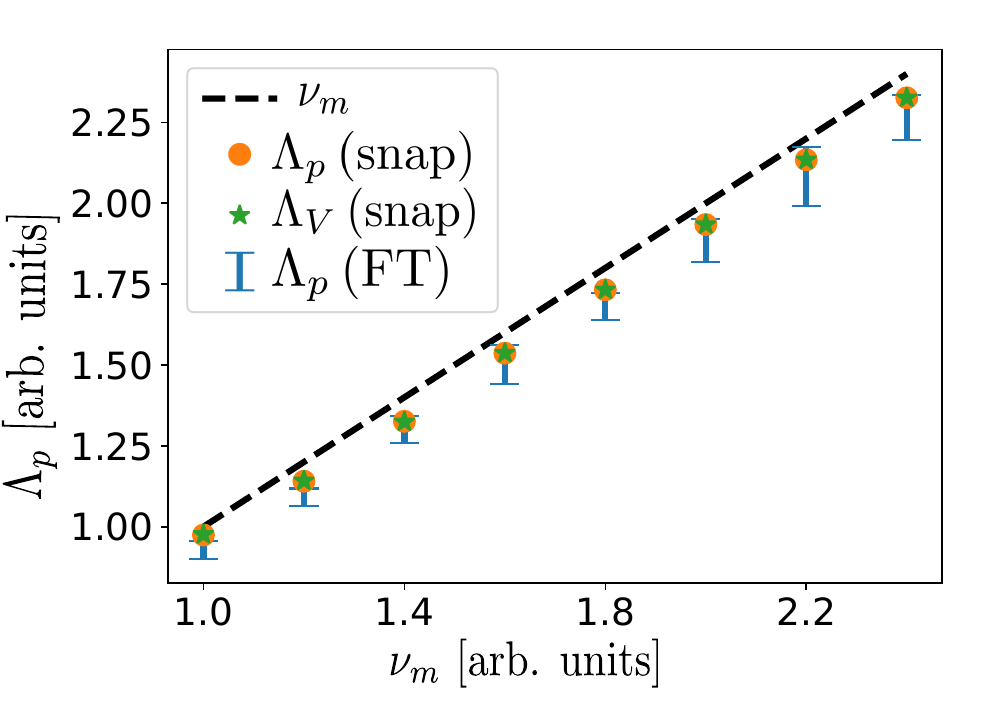}
  \caption{Population growth rate $\Lambda_P$
versus the mean single cell growth rate $\nu_m$: 
from a population snapshot (orange circles), from the growth rate of the total 
volume (green stars), and from Eq.~\eqref{FT_Lambda}.
Here the cell growth rate is 
taken from the normal distribution 
$\mathcal{N}(\nu_m,\sigma_\nu)$ and $B(x,\nu)=\nu x$. Error bars have 
been obtained by using the fluctuation relation on $1000$ trajectories and then repeating the 
estimation another 50 times.
}
\label{figlambda}
\end{figure}

Since the variability of single cell growth rate is known to be 
important experimentally \cite{Brenner2015}, 
we now discuss its role on the population growth rate in light of our results. 
A simple way to study this question in a simulation is to assume that the single cell
growth rate $\nu$ is distributed according to a normal distribution of mean $\nu_m$ and variance
$\sigma_\nu$. 
This is what we  have done in Fig.~\ref{figlambda}, where  
a division rate of the form $B(x,\nu)=\nu x$ has been used.
Although the values taken by $\nu$ are then uncorrelated 
from one division to the next, 
correlations between the mother and daugther generation times are still 
present due to the size dependence of the division rate.
This figure compares several determinations of 
the population growth rate $\Lambda_p$ as function of $\nu_m$.
In the absence of variability where $\sigma_\nu=0$, we have $\Lambda_p=\nu_m$, which is 
shown as a black dashed line in the figure. 
In the presence of variability, this figure confirms that 
the growth rate of the total volume $\Lambda_V$ equals 
the growth rate of the population where both of them 
have been measured from the statistics of the final population at a fixed time.
Importantly, such a determination of the population growth rate also agrees (within errors bars)
with the one based on Eq.~\eqref{FT_Lambda} using lineage trajectories. 
Therefore, this shows that Eq.~\eqref{FT_Lambda} could be used as a 
numerical method to determine a population growth rate based on lineage statistics.
 
Another striking feature of Fig.~\ref{figlambda} is that regardless of the 
determination of $\Lambda_p$, all the points are below the dashed line. 
The interpretation is that in a snapshot at time $t$,
it is less likely to see cells with a short generation time (corresponding 
to large single cell growth rates), therefore the distribution is biased towards 
small single cell growth rate \cite{Lin2017}. Since the population growth rate generally
increases with respect to the single cell growth rate $\nu_m$, this bias 
leads to a decrease of the population growth rate with respect to the case  
of no variability in the single cell growth rate.

As mentioned in the introduction, the fluctuation relation of Eq.~\eqref{FT} 
includes in itself a statistical biais: when 
choosing uniformly one individual in a population, an individual belonging to a lineage 
with prolific ancestors is more likely to be chosen, as a result, the jump rate on a lineage 
must be multiplied by the mean number of offsprings. 
Although variability in the single cell growth rate also introduces a form 
of statistical bias as explained above, the biais is not exactly the same one  
as that contained in the fluctuation relation. 
In any case, we would like to point out a comprehensive theoretical study on the effect 
of variability on the population growth rate, namely \cite{Olivier2017}.
This study confirms that in the case of size models with i.i.d. single cell 
growth rates, variability indeed lowers the Malthusian growth rate as observed in 
figure \ref{figlambda}. This work also discusses age models, with and without correlations 
in single cell growth rates, and concludes that in general, 
variability may lead to either a positive or negative trend on the population growth rate.

\subsection{Consequences for the distribution of generation times}
 
\begin{figure}[t]
\centering
\includegraphics[scale=0.35]{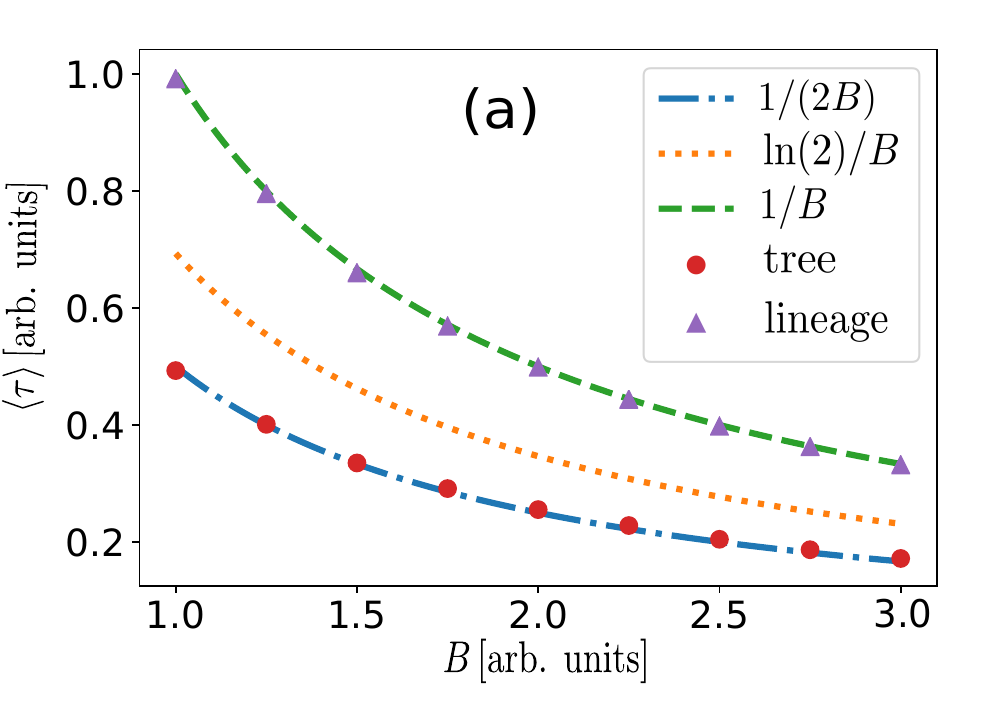}
\includegraphics[scale=0.35]{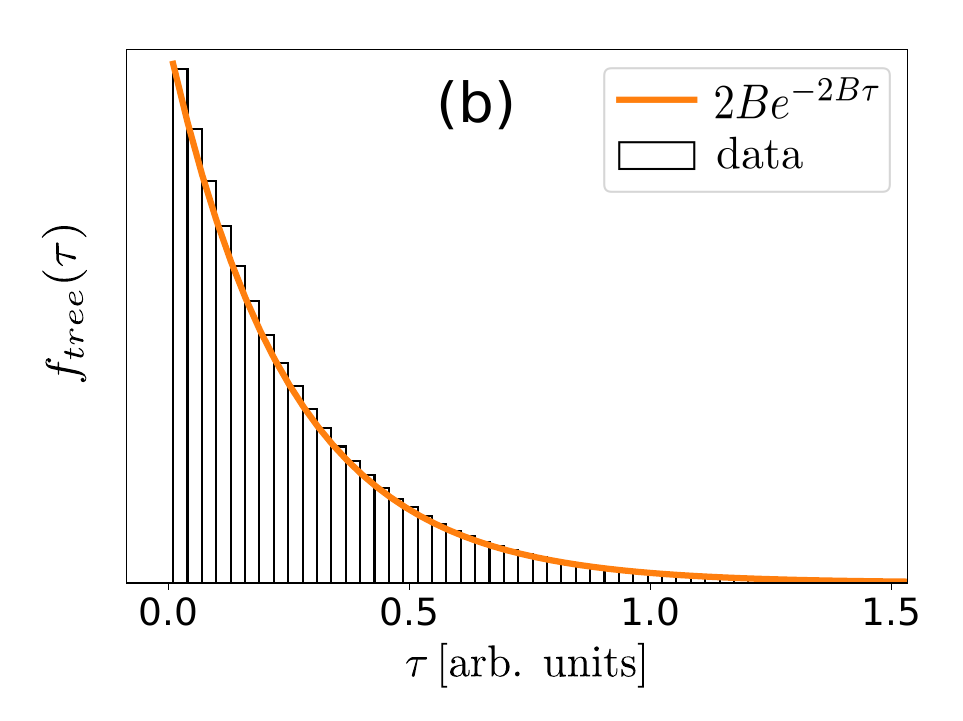}
  \caption{(a) Mean generation times evaluated in a tree (red circles) 
and in lineage (violet triangles) against the division rate $B$. 
Theoretical predictions are shown as dashed lines and 
the doubling time $T_d=\ln(2)/B$ is shown as a dotted line. 
(b) Distribution of generation times in the tree $f_{\T{tree}}(\tau)$. 
In these figures, the division 
rate, $B$ and the single cell growth rate, $\nu$, are constant and equal to each other.}
  \label{fig3}
\end{figure}

An important quantity in population dynamics is the distribution of 
generation times $f(\tau)$. This quantity can be evaluated from the observable \cite{Yuichi2011}:
\be
\label{AK}
A_K=  \frac{1}{K} \sum_{k=1}^K \delta(\tau-\tau_k),
\ee
where the index $k$ runs over all the $K$ cell cycles which have appeared 
in the trajectory that starts from $t=0$ to final time $t$. This observable can 
be evaluated either on a lineage or on a tree. 
By reporting $A_K$ as the observable $A$ in Eq.~\eqref{FT}, one deduces the relation 
\be
\label{rel-fpop}
f_{\T{tree},B}(\tau)= \langle \frac{1}{K} \sum_{k=1}^K \delta(\tau-\tau_k) e^{W_t - t\Lambda_t} 
\rangle_{\T{lin},2B},
\ee
where a summation over the random variable $K$ and a dependence on the final 
time $t$ are implicit. 
In the particular case where the division rate $B$ is constant,  
$W_t=t \Lambda_t$ and therefore $f_{\T{tree},B}(\tau)=f_{\T{lin},2B}(\tau)$. 
In this case, 
the generation time distribution in a lineage is the simple exponential
$f_{\T{lin},B}(\tau)=B \cdot \exp{(B\tau)}$ with mean $1/B$. 
It follows that $f_{\T{tree},B}(\tau)=2B \cdot \exp{(2B\tau)}$ with mean $1/(2B)$.
Fig.~\ref{fig3} confirms that the distribution of generation times
has the expected properties.

\section{A second fluctuation theorem to relate lineage and tree statistics}
\label{size-2}

When the division rate $B$ is not constant, the distribution of generation 
times will no longer be exponential, but we may still wonder how 
mean generation times observed at the lineage and tree 
levels compare to each other.
In order to address this issue, we derive a different fluctuation theorem 
that connects this time the lineage and tree statistics with the {\it same} division rate $B$.
More precisely, it follows from a direct comparison of \eqref{path-lin} and~\eqref{path-pop-2} taking
again $P_0=p_0$.
Since the division rate is the same in both probability distributions, we stick to the notations introduced above, except that now 
the index $B$ will be omitted. 

This allows to write
\begin{equation}
 \label{rel-1b}
 \ppo[\{x_k,\nu_k,t_k\}]=\pli[\{x_k,\nu_k,t_k\}]\exp\big[K\ln m-t\Lambda_t\big].
\end{equation}
Now, by multiplying the above relation by an arbitrary trajectory-observable $A$ and taking $m=2$, 
we obtain:
\be
\label{FT2}
\langle A( \{ \y \} ) \rangle_{\T{tree}}= \langle A( \{ \y \} )  
e^{K \ln 2-t \Lambda_t} \rangle_{\T{lin}},
\ee
where $K=K( \{ \y \} )$ counts as in Eq.~\eqref{AK} the number of divisions.

In the particular case 
where $A( \{ \y \} )=\delta ( \y - \y(t) ) \delta_{K, K(t)}$, 
Eq.~\eqref{FT2} leads upon averaging, 
to a relation between the joint probability distributions of size, growth rate and number of divisions
at the lineage and tree levels \cite{Nozoe2017}:
\begin{equation}
 \label{local-FT}
 P^{\T{\T{tree}}}(x,\nu,K)=2^K\,e^{-\Lambda_p\,t}P^{\T{\T{lin}}}(x,\nu,K),
\end{equation}
which we call a \emph{local} fluctuation relation.
Averages over lineages within a population can be carried out with respect  
to either a chronological or to a retrospective distribution 
\cite{Sughiyama2015, Kobayashi2015,Yuichi2011}, 
which correspond respectively to our lineage and tree probability distributions.
Let us briefly comment on a connection to a discussion presented 
in Ref.~\cite{Nozoe2017}. 
Elimination of $K$ in Eq.~\eqref{local-FT} leads to a fluctuation theorem only involving phenotypic traits $x$ and $\nu$:
\begin{align}
 \label{local-FT-1}
 P^{\T{\T{tree}}}(x,\nu) &=\sum_K\,P^{\T{\T{tree}}}(x,\nu,K)\nonumber\\
 &=e^{-\Lambda_p\,t}\sum_K\,2^K\,P^{\T{\T{lin}}}(x,\nu,K)\nonumber\\
 &=e^{-\Lambda_p\,t}P^{\T{\T{lin}}}(x,\nu)\sum_K\,2^K\,R^{\T{lin}}(K|x,\nu)\nonumber\\
 &\equiv e^{[h(x,\nu)-\Lambda_p]t}P^{\T{\T{lin}}}(x,\nu),
\end{align}
where we have introduced the probability of the number of division events conditioned on size and growth rate at the lineage level, 
$R^{\T{lin}}(K|x,\nu)$ and the equivalent of the ``fitness landscape'' of Ref.~\cite{Nozoe2017} reads
\begin{equation}
 \label{fitness}
 h(x,\nu)=\frac{1}{t}\ln\langle 2^K|x,\nu\rangle=\frac{1}{t}\ln\bigg(\sum_K\,2^K\,R^{\T{lin}}(K|x,\nu)\bigg).
\end{equation}
By summing over $K$ in Eq.~\eqref{local-FT}, one obtains
\be
 \label{local-FT-2}
 P^{\T{\T{tree}}}(x,\nu) = e^{[h(x,\nu)-\Lambda_p]t}P^{\T{\T{lin}}}(x,\nu),
\ee
in terms of a function $h(x,\nu)$ called ``fitness landscape'' 
in Ref.~\cite{Nozoe2017}.
Eqs.~\eqref{local-FT}-\eqref{local-FT-2} show that the knowledge of the two phenotypic 
probability distributions $P^{\T{\T{tree}}}$ and $P^{\T{\T{lin}}}$ can be used to infer a fitness function for size and growth rate.

\subsection{Inequalities for mean generation times}

Let us also introduce the Kullback-Leibler divergence 
between two probabilities $p$ and $q$:
\be
D(p|q)=\int dx \, p(x) \ln \frac{p(x)}{q(x)} \ge 0.
\ee
Using the fluctuation relation of Eq.~\eqref{rel-1b}, we obtain
\be
D(\pli | \ppo) = -\langle K \rangle_{\T{lin}} \ln 2 + t \Lambda_t.
\ee
On large times $t$, we can use the relation $\langle \tau \rangle_{\T{lin}}=t/ \langle
 K \rangle_{\T{lin}}$,
which together with the definition of the population doubling time $T_d=\ln2/\Lambda_t$,
 leads to the right inequality in 
\be
\label{inequalities}
\langle \tau \rangle_{\T{tree}} \le T_d \le \langle \tau \rangle_{\T{lin}},
\ee
while the left inequality follows 
very similarly 
using $D(\ppo | \pli)$.

\begin{figure}[ht]
\centering
\includegraphics[scale=0.35]{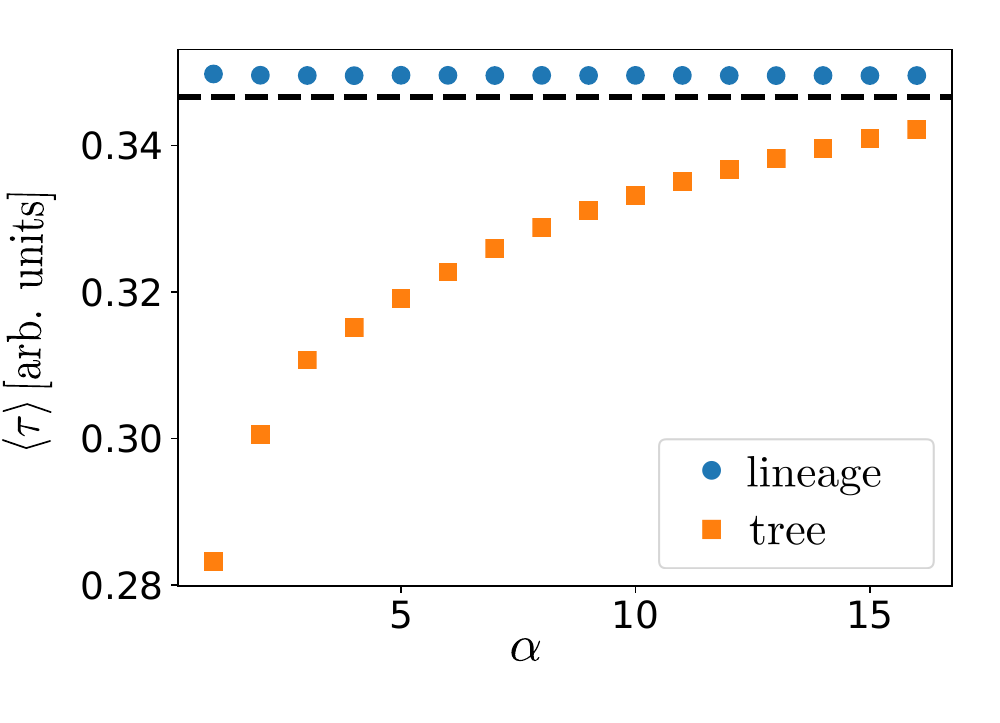}
  \caption{Mean generation times measured in a lineage or in a tree versus
the exponent $\alpha$ entering in the division rate $B(x,\nu)=\nu x^\alpha$.
The dashed line represents the doubling time $\ln2/\Lambda$, the single 
cell growth rate is $\nu=2$, and values $\alpha \in [1,16]$ are shown.}
  \label{fig4}
\end{figure}

In the case that $B$ is constant shown in Fig.~\ref{fig3}a, 
Eq.~\eqref{inequalities} is trivially satisfied. 
For $B$ non-constant of the form $\nu x^\alpha$, 
the inequalities are verified numerically in Fig.~\ref{fig4}.
This figure shows that the mean generation time for lineage (resp. tree) 
approaches the doubling time in the limit of large $\alpha$, because in this 
limit the distribution of generation times becomes peaked at $T_d$.

\subsection{Illustration of the inequalities based on experimental data}

In this section, we present an illustration of the above inequalities, namely Eq.~\eqref{inequalities},
using the experimental data of Hashimoto et al. \cite{Hashimotoa2016}.
In this experiment, the single-cell growth dynamics of a population of {\it E. coli} placed in 
a constant environment has been tracked in a flow cytometer. From an analysis of single 
cell lineages, this experiment yields measurements of the  
population doubling time, together with the two mean generation times discussed above. 
The terminology used in this paper is different from the one we have introduced here but
one can show that the distributions $g$ and $g^*$ introduced in that work 
correspond to what we call the lineage and tree distributions, respectively.

\begin{figure}[t]
\centering
  \includegraphics[scale=0.35]{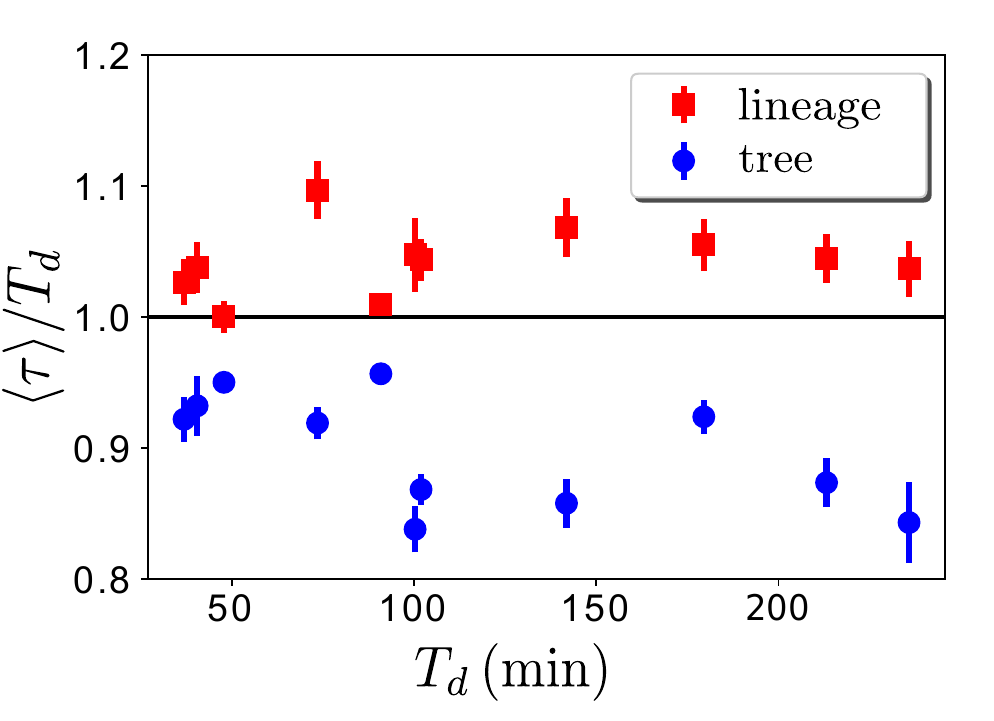}
  \caption{Mean generation time evaluated on a tree (blue filled circles)
and on a lineage (red filled squares) normalized by the population doubling time $T_d$
as a function of the population doubling time $T_d$. This data has been extracted from 
\cite{Hashimotoa2016}.}
  \label{fig5}
\end{figure}

In the figure 3B of the main text of \cite{Hashimotoa2016}, the mean generation 
time along a lineage is shown as a function of the population doubling time $T_d$ 
for various bacteria strains, 
which illustrates the right inequality of Eq.~\eqref{inequalities}. Fortunately, in 
the Suppl. Mat. of the paper, the data needed to plot the tree average is also given. 
For this reason, we show in Fig.~\ref{fig5}, the combined data which illustrates 
both inequalities of Eq.~\eqref{inequalities}.

This data has been interpreted in the framework of age-controlled model, 
assuming negligible mother-daughter correlations, which the authors have
checked with their data \cite{Hashimotoa2016}. 
They have also shown theoretically, as we also find in the next section devoted to age models, 
that Eq.~\eqref{inequalities} hold for 
age-controlled model assuming no mother-daughter correlations.
The important point we would like to make here is 
that this experimental data shown in Fig.~\ref{fig5} 
is also fully compatible with our predictions for 
size control models according to our Eq.~\eqref{inequalities}.
The strength of our derivation lies precisely in the fact that 
this result holds for a broad class of size-controlled models, 
without having to prescribe a precise form of the division rate or of the kernel $\Sigma$.

\section{Results for age-controlled models}
\label{sec:age}

\subsection{Definition of age-controlled models}

When the division rate depends on the age of the cells instead of their size, the structure of the model is 
rather different from that of the previous subsection. 
Let us now introduce a further distinction between two types of age models. In the first type, the interdivision times 
of mother and daughter cells are uncorrelated, and the division rate is determined 
by the age of the cells only. Such a model is usually termed independent generation times (IGT) model or 
Bellmann-Harris process \cite{Kimmel2015}. 
In a second type of models, the division rate may depend on 
other variables besides the age, and as a result, mother-daughter correlations will be present. 
Although many results for IGT models have already appeared in the literature, it is 
needed to go through them in order to understand what changes when correlations are present.

In the case of the IGT type of models, the density of cells having age $a$ in the population 
at time $t$, $n(a,t)$, satisfies the evolution equation
\begin{equation}
 \label{IGT-pop-den}
 \big(\partial_t+\partial_a\big)n(a,t)=-B(a)n(a,t),
\end{equation}
with the boundary condition:
\begin{equation}
 \label{BC-IGT-pop-den}
 n(0,t)=2\int_0^\infty\,B(a)n(a,t)da.
\end{equation}

As before, $B(a)$ denotes the age-dependent division rate. The physical interpretation of the boundary
 condition~\eqref{BC-IGT-pop-den} is clear: Each dividing cell gives rise to two newborn cells (i.e. two cells
with age $a=0$). The total number of cells in the population at time $t$ 
follows by integration of the density, $N(t)=\int\,n(a,t)da$.

As in the case of size control, lineage dynamics can be directly encoded in the evolution of the age distribution. Such dynamics reads
\begin{equation}
 \label{IGT-lin}
 \big(\partial_t+\partial_a\big)p(a,t)=-B(a)p(a,t).
\end{equation}
which is complemented by the boundary condition:
\begin{equation}
 \label{BC-IGT-lin}
 p(0,t)=\int_0^\infty\,B(a)p(a,t)da,
\end{equation}
so that probability is conserved and $p(a,t)$ is normalized.

In a second type of models, correlations in the inter-division times are accounted for by adding an extra 
dependence of the division rate on the growth rate, $B(a,\nu)$, while introducing at the same time
correlations between the growth rate of mother and daughter cells. The model then reads
\begin{equation}
 \label{corr-pop-den}
 \big(\partial_t+\partial_a\big)n(a,\nu,t)=-B(a,\nu)n(a,\nu,t),
\end{equation}
\begin{equation}
 \label{BC-corr-pop-den}
 n(0,\nu,t)=2\int_0^\infty\,da\int_0^\infty\,d\nu'\Sigma(\nu|\nu')B(a,\nu')n(a,\nu',t),
\end{equation}
at the population level, and
\begin{equation}
 \label{corr-lin}
 \big(\partial_t+\partial_a\big)p(a,\nu,t)=-B(a,\nu)p(a,\nu,t).
\end{equation}
\begin{equation}
 \label{BC-corr-lin}
 p(0,\nu,t)=\int_0^\infty\,da\int_0^\infty\,d\nu'\Sigma(\nu|\nu')B(a,\nu')p(a,\nu',t),
\end{equation}
at the lineage level.

\subsection{Generation time distribution}

Beyond cell size control models, one can also 
consider age models, which may have or not mother-daughter correlations. 
Let us first consider the case where correlations are absent, the so-called
IGT model, and let us focus on the distribution of generation times either in 
a lineage or in a population.

As in the case of size control, lineage dynamics of age-structured models can be directly encoded in the evolution of the age distribution, as prescribed by
Eqs.~\eqref{IGT-lin} and~\eqref{BC-IGT-lin}. Let us consider steady-state conditions:

\begin{equation}
 \label{IGT-lin-ss}
\partial_ap(a)=-B(a)p(a),
\end{equation}
\begin{equation}
 \label{BC-IGT-lin-ss}
 p(0)=\int_0^\infty\,B(a)p(a)da.
\end{equation}
A nice feature of age models is that the generation-time distribution can be accessed directly. 
This is so because generation time distribution is the age distribution of the dividing cells. 
We proceed to compute this distribution for individual lineages in age-structured IGT models. 
First, note that
from~\eqref{IGT-lin-ss} immediately follows that
\begin{equation}
 \label{pss}
 p(a)=p(0)\exp\bigg[-\int_0^a\,B(a')da'\bigg].
\end{equation}
Relying on the relation between generation time distribution and age distribution of \emph{dividing} cells, 
we can write 
\begin{align}
 \label{gen-t-IGT-lin}
 f_{\T{\T{lin}}}(\tau) &=\frac{B(\tau)p(\tau)}{\int_0^\infty\,B(a)p(a)da}\nonumber\\
 &\equiv\,B(\tau)\exp\bigg[-\int_0^\tau\,B(a)da\bigg],
\end{align}
where we have used~\eqref{BC-IGT-lin-ss} and~\eqref{pss}.

Now in order to obtain the distribution of generation times at the population level,
we start from Eqs.~\eqref{IGT-pop-den} and~\eqref{BC-IGT-pop-den}. Again, we focus on stationary
 conditions for which the total number of cells in the population 
grows exponentially, as $N(t)=e^{\Lambda_p t}$. In that case, de density can be written in terms
of the stationary probability density of cells with a given age as $n(a,t)=e^{\Lambda_p t}P(a)$,
 where $P(a)$ is the stationary age distribution of the population. We have:
\begin{equation}
 \label{IGT-pop-ss}
 \partial_a P(a)=-\big[\Lambda_p+B(a)\big]P(a),
\end{equation}
with the boundary condition:
\begin{equation}
 \label{BC-IGT-pop-ss}
 P(0)=2\int_0^\infty\,B(a)P(a)da.
\end{equation}
It is worth noting that normalization of $P(a)$ in~\eqref{IGT-pop-ss}, leads, using~\eqref{BC-IGT-pop-ss}, to the following identity
\begin{equation}
 \label{identity-IGT}
 \Lambda_p=\int_0^\infty\,B(a)P(a)da\equiv\frac{1}{2}P(0).
\end{equation}
We can now proceed to compute the generation time distribution, by computing the age distribution of dividing cells. We have first for the stationary distribution from~\eqref{IGT-pop-ss}:
\begin{align}
 \label{IGT-pop-ss-prob}
 P(a) &=P(0)\exp\bigg[-\Lambda_p a -\int_0^a\,B(a')da'\bigg]\nonumber\\
 &\equiv 2\Lambda_p\exp\bigg[-\Lambda_p a -\int_0^a\,B(a')da'\bigg],
\end{align}
where we have also used~\eqref{identity-IGT}. On passing by, we highlight an important relation for IGT models obtained from the normalization of $P(a)$ in Eq.~\eqref{IGT-pop-ss-prob}:
\begin{equation}
 \label{identidad}
 \int_0^\infty\,\exp\bigg[-\Lambda_p a -\int_0^a\,B(a')da'\bigg]da=\frac{1}{2\Lambda_p}.
\end{equation}

We can now calculate the generation time distribution, which reads
\begin{align}
 \label{gen-t-IGT-pop}
 f_{\T{\T{tree}}}(\tau) &=\frac{B(\tau)P(\tau)}{\int_0^\infty\,B(a)P(a)da}\nonumber\\
 &=2B(\tau)\exp\bigg[-\Lambda_p \tau -\int_0^\tau\,B(a)da\bigg].
\end{align}
Reading now from the result for the lineage, Eq.~\eqref{gen-t-IGT-lin}, we obtain :
\begin{equation}
 \label{DFT-IGT}
 f_{\T{\T{tree}}}(\tau)=2f_{\T{\T{lin}}}(\tau)e^{-\Lambda_p \tau},
\end{equation}
which corresponds to the result derived in Ref~\cite{Hashimotoa2016} 
with the identification of their generation time distribution $g$ (resp. $g^*$) 
with our distributions $f_{\T{lin}}$ (resp. $f_{\T{tree}}$).

Using Eq.~\eqref{DFT-IGT} we have, for instance:
\begin{align}
 \label{IGT-ineq-1}
 D(f_{\T{tree}}||f_{\T{\T{lin}}}) &=\int_0^\infty\,f_{\T{tree}}(\tau)\ln\frac{f_{\T{tree}}(\tau)}{f_{\T{\T{lin}}}(\tau)}\,d\tau\nonumber\\
 &=\ln2\bigg[1-\frac{\langle\tau\rangle_{\T{tree}}}{T_d}\bigg]\ge0\nonumber\\
 &\Rightarrow\,\langle\tau\rangle_{\T{tree}}\leq T_d,
\end{align}
where as usual the population doubling time reads $T_d=\ln2/\Lambda_p$. It is straightforward to 
prove the second inequality using the same technique. We then conclude that for IGT models, 
one has the same result as obtained for size structured populations in 
Eq.~\eqref{inequalities}, i.e., 
\begin{equation}
 \label{fin-IGT}
 \langle\tau\rangle_{\T{tree}}\leq T_d\leq\langle\tau\rangle_{\T{lin}}.
\end{equation}

\subsection{Beyond uncorrelated age models}

In view of the result of previous section, it is then natural to ask what happens in 
the more complex case in which mother-daughter correlations are present.
In appendix \ref{app:corr}, we derived a generalization of Eq.~\eqref{DFT-IGT}
for that case, namely:
\begin{equation}
 \label{DFT-corr}
 f_{\T{\T{tree}}}(\tau,\nu)=2\frac{\rt(\nu)}{\rl(\nu)}f_{\T{\T{lin}}}(\tau,\nu)e^{-\Lambda_p \tau},
\end{equation}
where $\rl(\nu)$ (resp. $\rt(\nu)$) represent 
the growth rate distributions of newborn cells at the lineage (resp. tree level).
The presence of these two new probability distributions is entirely due to mother-daughter 
correlations. As a result, the inequalities \eqref{inequalities} (identical to \eqref{fin-IGT}) 
do not necessarily hold 
for age models with correlations. An example where they are indeed violated  
can be found in the model with correlated generation times studied 
in Ref.~\cite{Amir2018} in some range of parameters.

\section{Conclusion}
\label{conc}

In conclusion, we have established several fluctuation relations which 
relate observables measured at the lineage and tree levels.
We have deduced from the second fluctuation relation that 
mean generation times in a lineage are larger 
than the population doubling time,
while mean generation times in a tree are smaller than
the population doubling time.
We have found that the experimental data of Hashimoto et al. fully confirm  
this observation  \cite{Hashimotoa2016}.
We conclude from this that this data is compatible either with the uncorrelated 
age-models used by these authors to analyze their data, or 
with the class of cell size control mechanisms considered in this paper.

Our approach being general, it could be extended to cover more complex cases such 
asymmetric divisions relevant for yeast cells, non-exponential regimes of growth, 
relevant for eukariots and other mechanisms of cell aging \cite{Stewart2005}. 
While we have mainly focused on the control of the size variable,  
extension of this formalism to other variables not directly linked to cell size is possible, 
one choice being for instance the protein copy numbers \cite{Brenner2015,Thomas2017}.
 
We also find that the variability of single cell growth has a negative 
impact on the population growth rate in the absence of mother-daughter growth-rate 
correlations when the division rate is $B(x,\nu)=\nu x$. A positive impact 
due to correlations in the inter-division times has been reported in some other study \cite{Amir2018}, while more 
generally a positive or negative impact should be expected depending on the form 
of the division rate \cite{Olivier2017}. 
All these recent results suggest that generation times are under a strong evolutionary 
pressure in which single cell variability and correlations over generations \cite{Rivoire2014} 
play an important role.

In the future, we would like to study systems where the division rate is
controlled simultaneosly by the size and the age of the cell, 
which represents a situation of major biological relevance~\cite{Osella201313715}.
Finally, while this work was under review, two new studies of cell growth dynamics  
have appeared, which relate either to our pathwise formulation \cite{Sughiyama2018} 
or to our analysis of generation time distributions \cite{Jafarpour2018}.

\begin{acknowledgments}
R.G.G. was supported by the Agence Nationale de la Recherche 
(ANR-16-CE11-0026-03) and by Labex CelTisPhysBio (ANR-10-LBX-0038).
We would like to thank E. Braun, L. Robert, A. Olivier, T. J. Kobayashi and Y. Wakamoto
for stimulating discussions.
\end{acknowledgments}

\appendix

\begin{widetext}

\section{Path integral representation of the dynamics for size-controlled models}
\label{app:path}

\subsection{Population level}

Let us start by building a path integral representation associated to the 
evolution of the number density of cells in the population case, Eq.~\eqref{CRMpopulation2}. 
Here, we will allow for an arbitrary number of offsprings $m$ for generality, although only $m=2$
was considered above. We emphasize that $m$ should be independent of the state of the system.
Let us treat the following term 
\be
\label{def_f}
f(\y,t)= m\int d\y'\Sigma(\y|\y')B(\y')n(\y',t),
\ee
in Eq.~\eqref{CRMpopulation2} as a perturbation. 
The \emph{growth propagator} $G_B$ of the unperturbed dynamics is such that 
\begin{equation}
 \label{weak}
 \partial_tG_B(x,\nu,t|x',t') =-\nu\partial_x\big[xG_B(x,\nu,t|x',t')\big]-B(x,\nu)G_B(x,\nu,t|x',t'),
\end{equation}
with initial condition $G_B(x,\nu,t'|x',t')=\delta(x-x')$.
Then using these equations, one can check that
\begin{equation}
 \label{gen-identity}
n(x,\nu,t)=\int_0^\infty dx_0\,G_B(x,\nu, t|x_0,0)n_0(x_0,\nu)+\int_0^t dt'\,\int_0^\infty dx'\,G_B(x,\nu,t|x',t')f(x',\nu,t'),
\end{equation}
is equivalent to the initial problem given in Eq.~\eqref{CRMpopulation2}.
By explicitly using the definition of $f$ from Eq.~\eqref{def_f}, one obtains
\begin{align}
 \label{gen-identity-1}
 n(x,\nu,t) &=\int_0^\infty dx_0\,G_B(x,\nu, t|x_0,0)n_0(x_0,\nu)+ m \int_0^t dt_1\,\int_0^\infty dx_1\,G_B(x,\nu,t|x_1,t_1)\times\nonumber\\
 &\times\int_0^\infty d\nu_0\int_0^\infty dz\;\Sigma(x_1,\nu,|z,\nu_0)B(z,\nu_0)n(z,\nu_0,t_1),
 \end{align}
which allows to find an explicit solution for $n(x,\nu,t)$ iteratively. 

The explicit solution of Eq.~\eqref{weak} is
\begin{equation}
 \label{GB}
 G_B(x,\nu,t|x',t')=\delta\big(x-x'e^{\nu(t-t')}\big)\exp\bigg[-\int_{t'}^t\,d\tau B\big(x'e^{\nu(\tau-t')},\,\nu\big)\bigg].
\end{equation}
which allows us to write: 
 \begin{align}
 \label{gen-identity-2}
 n(x,\nu,t) &=\int_0^\infty dx_0\,\delta\big(x-x_0e^{\nu t}\big)\exp\bigg[-\int_{0}^t\,d\tau B\big(x_0e^{\nu\tau},\,\nu\big)\bigg]n_0(x_0,\nu)\;+\nonumber\\
 &+m\int_0^t dt_1\,\int_0^\infty d\nu_0\int_0^\infty dx_1\int_0^\infty dx_0\,\delta\big(x-x_1e^{\nu(t-t_1)}\big)\exp\bigg[-\int_{t_1}^t\,d\tau B\big(x_1e^{\nu(\tau-t_1)},\,\nu\big)\bigg]\times\nonumber\\
 &\times\Sigma\big(x_1,\nu,\big|x_0e^{\nu_0 t_1},\nu_0\big)B\big(x_0e^{\nu_0 t_1},\nu_0\big)\exp\bigg[-\int_{0}^{t_1}\,d\tau B\big(x_0e^{\nu_0\tau},\,\nu_0\big)\bigg]n_0(x_0,\nu_0)\;+\ldots
\end{align}
or more compactly:
\begin{align}
 \label{path-pop}
n(x,\nu,t) &=\sum_{K=0}^\infty m^K\int_0^t dt_K\ldots\int_0^{t_2} dt_1\int_0^{\infty}\prod_{k=0}^K dx_k\,d\nu_k\,\delta(\nu-\nu_K)\delta\big(x-x_Ke^{\nu_K(t-t_K)}\big)\,
n_0(x_0,\nu_0)\times\nonumber\\
 &\times\exp\bigg[-\int_{0}^{t} d\tau B\big(x(\tau),\nu(\tau)\big)\bigg]
 \prod_{k=1}^K\trans\bigg(x_k,\nu_k,\big|x_{k-1}e^{\nu_{k-1}( t_k-t_{k-1})},\nu_{k-1}\bigg),
\end{align}
where trajectories explicity appearing in the exponential in the r.h.s. of~\eqref{path-pop} are given as $\nu(\tau)=\nu_k$, and $x(\tau)=x_k\exp(\nu_k(\tau-t_k))$, for $\tau\in(t_k, t_{k+1}]$, while $k=0,1,\ldots,K$.
In our notations, $t_0=0$ and $t_{K+1}=t$. In addition, the transition matrix is given as $\trans(x,\nu|x',\nu')=\Sigma(x,\nu|x',\nu')B(x',\nu')$.

The last step now consists in noticing that the object propagating trajectories from 
$t_0=0$ up to time $t$ in~\eqref{path-pop} is not yet a path probability because it is not properly normalized. 
To deal with
this issue it is good to pass from number densisites to population-level probability densities, 
$P(x,\nu,t)=N(t)^{-1}n(x,\nu,t)$, and $P_0(x,\nu)=N(0)^{-1}n_0(x,\nu)$. We can now write in terms of these quantities:
\begin{equation}
 \label{path-pop-1}
 P(x,\nu,t) =\sum_{K=0}^\infty\int_0^t dt_K\ldots\int_0^{t_2} dt_1\int_0^{\infty}\prod_{k=0}^K \,d\nu_k\,\int_0^{\infty}\prod_{k=0}^{K+1}\,dx_k\, \delta(\nu-\nu_K)\delta(x-x_{K+1})\ppop[\{x_k,\nu_k,t_k\}],
\end{equation}
where the object
\begin{align}
 \label{path-pop-2}
 \ppop[\{x_k,\nu_k,t_k\}] &= m^K\,\delta\big(x_{K+1}-x_Ke^{\nu_K(t-t_K)}\big)\exp\bigg[-t\Lambda_t-\int_{0}^{t} d\tau B\big(x(\tau),\nu(\tau)\big)\bigg]\times\nonumber\\
 &\times\prod_{k=1}^K\trans\bigg(x_k,\nu_k,\big|x_{k-1}e^{\nu_{k-1}( t_k-t_{k-1})},\nu_{k-1}\bigg)P_0(x_0,\nu_0),
\end{align}
\end{widetext}
is now properly normalized and can be identified with the correct path propability 
generating averages of all observables related to the number density at the population level. 
We have added a subscript $B$ to indicate  that the division rate is given by $B(x,\nu)$. 
This will be important later in the derivation of fluctuation theorems.
Note that when passing from densities to probability densities, 
a new term has appeared in the argument of the exponential namely $\Lambda_t$, which is connected to the population 
growth rate, $\Lambda_p$, by Eq.~\eqref{GR}.

\subsection{Lineage level}

The starting point to derive the path probability for lineage observables is the evolution equation 
for the probability density of size and growth rate, Eq.~\eqref{CRMlineage2}
Except for the absence of the factor two in front of the integral, the structure of the equations are the same and the derivation follows along exactly as in the population case.
We provide the final result:
\begin{widetext}
\begin{align}
 \label{path-lin}
 \plin[\{x_k,\nu_k,t_k\}] &= \,\delta\big(x_{K+1}-x_Ke^{\nu_K(t-t_K)}\big)\exp\bigg[-\int_{0}^{t} d\tau B\big(x(\tau),\nu(\tau)\big)\bigg]\times\nonumber\\
 &\times\prod_{k=1}^K\trans\bigg(x_k,\nu_k,\big|x_{k-1}e^{\nu_{k-1}( t_k-t_{k-1})},\nu_{k-1}\bigg)p_0(x_0,\nu_0),
\end{align}
which can be readily shown to be properly normalized. Here $p_0$ is the distribution of initial conditions for the 
lineage. Note that we have introduced $p_0$, which could be different from the $P_0$ introduced earlier 
as the initial condition of the population.

\subsection{Derivation of fluctuation relations}
\label{app:FT-size}

We can now compare 
path probabilities representations at the population and lineage levels given by
\eqref{path-lin} and~\eqref{path-pop-2} with each other.
We see that a possible way to bring both distributions 
``closer'' together, is to multiply the division rate at the lineage level by the factor $m$, 
and to consider 
a lineage starting from the same initial condition as that of the population. 
A possible choice of this initial condition consists, for instance, 
in considering a population dynamics starting from a single cell. 

In that case we have:
\begin{align}
 \label{path-lin-2B}
 \plim[\{x_k,\nu_k,t_k\}] &= m^K\,\delta\big(x_{K+1}-x_Ke^{\nu_K(t-t_K)}\big)\exp\bigg[-m\int_{0}^{t} 
d\tau B\big(x(\tau),\nu(\tau)\big)\bigg]\times\nonumber\\
 &\times\prod_{k=1}^K\trans\bigg(x_k,\nu_k,\big|x_{k-1}e^{\nu_{k-1}( t_k-t_{k-1})},\nu_{k-1}\bigg)P_0(x_0,\nu_0).
\end{align}
Then, the following relation holds from direct comparison of~\eqref{path-lin-2B} and~\eqref{path-pop-2}:
\begin{equation}
 \label{rel-1}
 \ppop[\{x_k,\nu_k,t_k\}]=\plim[\{x_k,\nu_k,t_k\}]\exp\bigg[ (m-1) \int_{0}^{t} d\tau B\big(x(\tau),\nu(\tau)\big)-t\Lambda_t\bigg].
\end{equation}

\section{Fluctuation theorem for correlated age models}
\label{app:corr}
\subsection{Lineage dynamics}

We now consider models in which interdivision times are correlated. The natural way in which these correlations arise is by inter-cell-cycle growth-rate fluctuations,
as given by Eqs.~\eqref{corr-lin} and~\eqref{BC-corr-lin}.
Growth-rate correlations are encoded in $\Sigma$, which is a properly normalized conditional probability. Again, we will focus on stationary conditions. It is simple to see from~\eqref{corr-lin} that
one can formally write the stationary distribution as:
\begin{equation}
 \label{corr-lin-ss}
 p(a,\nu)=p(0,\nu)\exp\bigg[-\int_0^a\,B(a',\nu)da'\bigg].
\end{equation}
To determine $p(0,\nu)$, we use~\eqref{BC-corr-lin} and~\eqref{corr-lin-ss}:
\begin{align}
 \label{g-lin}
 p(0,\nu) &=\int_0^\infty\,da\int_0^\infty\,d\nu'\Sigma(\nu|\nu')B(a,\nu')p(0,\nu')\exp\bigg[-\int_0^a\,B(a',\nu')da'\bigg]\nonumber\\
 &=-\int_0^\infty\,d\nu'\Sigma(\nu|\nu')p(0,\nu')\int_0^\infty\,da\,\frac{d}{da}\exp\bigg[-\int_0^a\,B(a',\nu')da'\bigg],
\end{align}

from where we get the following integral equation:
\begin{equation}
 \label{g-lin-1}
 p(0,\nu)=\int_0^\infty\,\Sigma(\nu|\nu')p(0,\nu')\,d\nu'.
\end{equation}
The generation time distribution can now be determined, again, as the age-distribution of dividing cells. 
It is worth considering  slightly more general object, i.e., the \emph{joint} probability distribution of interdivision time and growth rate
\begin{equation}
 \label{gen-t-corr-lin}
 f_{\T{\T{lin}}}(\tau,\nu) =\frac{B(\tau,\nu)p(\tau,\nu)}{\int_0^\infty\,da\int_0^\infty\,d\nu\,B(a,\nu)p(a,\nu)}
 =\frac{B(\tau,\nu)p(0,\nu)\exp\bigg[-\int_0^\tau\,B(a,\nu)da\bigg]}{\int_0^\infty\,p(0,\nu')\,d\nu'}.
\end{equation}
\end{widetext}
This result can be written in a more illuminating way by noticing that the growth rate distribution of newborn cells can be identified as
\begin{equation}
 \label{nb-lin}
 \rl(\nu)=\frac{p(0,\nu)}{\int_0^\infty\,p(0,\nu')\,d\nu'}.
\end{equation}
Furthermore, due to the linearity of Eq.~\eqref{g-lin-1}, and the fact that $\rl(\nu)$ differs from $p(0,\nu)$ only in a multiplicative constant, we have that $\rl(\nu)$ satisfies
\begin{equation}
 \label{nb-lin-1}
 \rl(\nu)=\int_0^\infty\,\Sigma(\nu|\nu')\rl(\nu')\,d\nu'.
\end{equation}
These observations then lead to the final result:
\begin{equation}
 \label{gen-t-corr-lin-joint}
 f_{\T{\T{lin}}}(\tau,\nu)=\rl(\nu)B(\tau,\nu)\exp\bigg[-\int_0^\tau\,B(a,\nu)da\bigg].
\end{equation}

\subsection{Population dynamics}

Let us now consider the population level. The stationary equation for the population age distribution reads
\begin{equation}
 \label{corr-pop}
 \partial_aP(a,\nu)=-\big[\Lambda_p+B(a,\nu)]P(a,\nu),
\end{equation}
with boundary condition
\begin{equation}
 \label{BC-corr-pop}
 P(0,\nu)=2\int_0^\infty\,da\int_0^\infty\,d\nu'\Sigma(\nu|\nu')B(a,\nu')P(a,\nu').
\end{equation}
We then have
\begin{equation}
 \label{corr-pop-ss}
 P(a,\nu)=P(0,\nu)\exp\bigg[-\Lambda_p a-\int_0^a\,B(a',\nu)da'\bigg],
\end{equation}
 Note that the normalization of $P$ gives the following condition:
 \begin{widetext}
\begin{align}
 \label{corr-norm-pop}
 \Lambda_p &=\int_0^\infty\,da\int_0^\infty\,d\nu\,B(a,\nu)P(a,\nu)=\int_0^\infty\,d\nu\,P(0,\nu)\int_0^\infty\,da\,B(a,\nu)\exp\bigg[-\Lambda_p a-\int_0^a\,B(a',\nu)da'\bigg]\nonumber\\
 &=-\int_0^\infty\,d\nu\,P(0,\nu)\int_0^\infty\,da\bigg(\frac{d}{da}+\Lambda_p\bigg)\exp\bigg[-\Lambda_p a-\int_0^a\,B(a',\nu)da'\bigg]\nonumber\\
 &=\int_0^\infty\,d\nu\,P(0,\nu)-\Lambda_p\int_0^\infty\,d\nu\,P(0,\nu)\Gamma(\nu),
\end{align}
where we have used~\eqref{corr-pop-ss} and introduced the function
\begin{equation}
\label{Gamma}
 \Gamma(\nu)=\int_0^\infty\,da\exp\bigg[-\Lambda_p a-\int_0^a\,B(a',\nu)da'\bigg].
\end{equation}
We can thus write for the growth rate of the population:
\begin{equation}
 \label{corr-pop-growth-rate}
 \Lambda_p=\frac{\int_0^\infty\,P(0,\nu)\,d\nu}{1+\int_0^\infty\,P(0,\nu)\Gamma(\nu)\,d\nu}.
\end{equation}
On the other hand, integrating directly in~\eqref{corr-pop-ss}, we get
\begin{equation}
 \label{corr-pop-growth-rate-1}
 1=\int_0^\infty\,da\int_0^\infty\,d\nu\,P(0,\nu)\exp\bigg[-\Lambda_p a-\int_0^a\,B(a',\nu)da'\bigg]=\int_0^\infty\,P(0,\nu)\Gamma(\nu)\,d\nu,
\end{equation}
so we have
\begin{equation}
 \label{corr-pop-growth-rate-2}
 \Lambda_p=\frac{1}{2}\int_0^\infty\,P(0,\nu)\,d\nu.
\end{equation}

As before, we can find an equation for $P(0,\nu)$ using~\eqref{BC-corr-pop} and the solution for $P(a,\nu)$, Eq.~\eqref{corr-pop-ss}:
\begin{align}
 \label{h-pop}
 P(0,\nu) &= 2\int_0^\infty\,da\int_0^\infty\,d\nu'\Sigma(\nu|\nu')P(0,\nu')\,B(a,\nu')\exp\bigg[-\Lambda_p a-\int_0^a\,B(a',\nu')da'\bigg]\nonumber\\
 &=-2\int_0^\infty\,d\nu'\Sigma(\nu|\nu')P(0,\nu')\int_0^\infty\,da\bigg(\frac{d}{da}+\Lambda_p\bigg)\exp\bigg[-\Lambda_p a-\int_0^a\,B(a',\nu')da'\bigg]\nonumber\\
 &=2\int_0^\infty\,d\nu'\Sigma(\nu|\nu')P(0,\nu')-2\Lambda_p\int_0^\infty\,d\nu'\Sigma(\nu|\nu')P(0,\nu')\int_0^\infty\,da\exp\bigg[-\Lambda_p a-\int_0^a\,B(a',\nu')da'\bigg],
\end{align}
so, we then have

\begin{equation}
 \label{h-pop-1}
 P(0,\nu)=2\int_0^\infty\,d\nu'\Sigma(\nu|\nu')\big[1-\Lambda_p\Gamma(\nu')\big]P(0,\nu').
\end{equation}

Let us now write the joint probability  distribution of interdivision times and single-cell growth rate:
\begin{equation}
 \label{gen-t-corr-pop-joint}
 f_{\T{\T{tree}}}(\tau,\nu)=\frac{B(\tau,\nu)P(\tau,\nu)}{\int_0^\infty\,da\int_0^\infty\,d\nu\,B(a,\nu)P(a,\nu)}
 =\frac{P(0,\nu)}{\Lambda_p}B(\tau,\nu)\exp\bigg[-\Lambda_p \tau-\int_0^\tau\,B(a,\nu)da\bigg].
\end{equation}
\end{widetext}
The condition~\eqref{corr-pop-growth-rate-2} implies that $P(0,\nu)/\Lambda_p=2\rt(\nu)$, where
\begin{equation}
\label{nb-tree}
\rt(\nu)=\frac{P(0,\nu)}{\int_0^\infty\,P(0,\nu)\,d\nu}
\end{equation}
can be identified, as we did in the lineage case, with the growth rate distribution of newborn cells, now at the tree level. We then have:
\begin{align}
 \label{gen-t-corr-pop-joint-1}
 f_{\T{\T{tree}}}(\tau,\nu) &=2\rt(\nu)B(\tau,\nu)\times\nonumber\\
 &\exp\bigg[-\Lambda_p \tau-\int_0^\tau\,B(a,\nu)da\bigg].
\end{align}
Note once more that the linearity of Eq.~\eqref{h-pop-1} and the fact that $P(0,\nu)$ and $\rt(\nu)$ differ only on a multiplicative factor, lead to the equation satisfied by $\rt$:
\begin{equation}
 \label{nb-tree-1}
 \rt(\nu)=2\int_0^\infty\,d\nu'\Sigma(\nu|\nu')\big[1-\Lambda_p\Gamma(\nu')\big]\rt(\nu').
\end{equation}

If we now compare~\eqref{gen-t-corr-pop-joint-1} and~\eqref{gen-t-corr-lin-joint}, we readily obtain Eq.~\eqref{DFT-corr}.
Before closing this paragraph some comments are in order. First, note that as Eqs.~\eqref{nb-lin-1} and~\eqref{nb-tree-1} are clearly different, one has $\rl(\nu)\neq\rt(\nu)$.
Nevertheless, in absence of fluctuations, when $\Sigma(\nu|\nu')=\delta(\nu-\nu')$, we have $\rl(\nu)=\rt(\nu)$. To illustrate this, let us consider, for instance, a population
starting from a single cell with growth rate $\nu_0$. As the growth rate remains the same in all cell cycles, we have $\rl(\nu)=\rt(\nu)\equiv\delta(\nu-\nu_0)$ at all times. Then, 
Eq.~\eqref{nb-lin-1} becomes tautological, while Eq.~\eqref{nb-tree-1} leads to the identity
\begin{equation}
 \label{identidad-1}
 2\big[1-\Lambda_p\Gamma(\nu_0)\big]=1,
\end{equation}
which is precisely the relation~\eqref{identidad} found for IGT models (recall the definition of $\Gamma$,~\eqref{Gamma}).

\subsection{Inequalities in correlated age models}

Let us now analyze the consequences of the generalized relation~\eqref{DFT-corr} 
for the inequalities. We have, for instance:
\begin{align}
 \label{corr-ineq-1}
 D(f_{\T{\T{tree}}}||f_{\T{\T{lin}}}) &=\int\,f_{\T{\T{tree}}}(\tau,\nu)\ln\frac{f_{\T{\T{tree}}}(\tau,\nu)}{f_{\T{\T{lin}}}(\tau,\nu)}\,d\tau d\nu\nonumber\\
 &=\ln2\bigg[1-\frac{\langle\tau\rangle_{\T{\T{tree}}}}{T_d}\bigg]+\nonumber\\
 &+\int_0^\infty \tl f_{\T{\T{tree}}}(\nu)\ln\frac{\rt(\nu)}{\rl(\nu)}\,d\nu\ge0,
\end{align}
where $\tl f_{\T{\T{tree}}}(\nu)=\int\,d\tau f_{\T{\T{tree}}}(\tau,\nu)$ is the marginal distribution of the growth rate of the \emph{dividing} cells. This result implies, in particular, that
\begin{equation}
 \label{corr-ineq-2}
 T_d-\langle\tau\rangle_{\T{\T{tree}}}\ge-\frac{T_d}{\ln2}\int_0^\infty \tl f_{\T{\T{tree}}}(\nu)\ln\frac{\rt(\nu)}{\rl(\nu)}\,d\nu.
\end{equation}

Given that the quantity in the right hand side of~\eqref{corr-ineq-2} does not have a definite sign (in particular, it is not necessarily positive), in this case the left inequality
in~\eqref{fin-IGT} (and Eq.~\eqref{inequalities}) may be violated. Repeating a similar argument, on arrives to the same conclusion for the right inequality.

\bibliographystyle{apsrev4-1}

\end{document}